%% file: tmi.tex
\documentclass[journal,twoside]{ieeecolor}
\usepackage{tmi}
\usepackage{cite}
\usepackage{amsmath,amssymb,amsfonts}
\usepackage{algorithmic}
\usepackage{graphicx}
\usepackage{textcomp}

\usepackage{multirow}
\usepackage{hyperref}
\usepackage{caption}

\def\BibTeX{{\rm B\kern-.05em{\sc i\kern-.025em b}\kern-.08em
    T\kern-.1667em\lower.7ex\hbox{E}\kern-.125emX}}
\newcommand{\hh}{\hspace{1pt}} 
\newcommand{\hf}{\hspace{4pt}} 
\newcommand{\he}{\hspace{2ex}}

\usepackage[normalem]{ulem}

\begin{document}

\title{Improving anatomical plausibility in medical image segmentation via hybrid graph neural networks: applications to chest x-ray analysis}

\author{Nicolás Gaggion, Lucas Mansilla, Candelaria Mosquera, Diego H. Milone and Enzo Ferrante
\thanks{N. Gaggion, L. Mansilla, D.H. Milone and E. Ferrante are with the Institute for Signals, Systems and Computational Intelligence, sinc(i) CONICET-UNL, Santa Fe, Argentina. (e-mails: ngaggion@sinc.unl.edu.ar, lmansilla@sinc.unl.edu.ar, dmilone@sinc.unl.edu.ar, eferrante@sinc.unl.edu.ar).}
\thanks{C. Mosquera is with the Health Informatics Department at Hospital Italiano de Buenos Aires and with Universidad Tecnológica Nacional, Buenos Aires, Argentina (e-mail: candelaria.mosquera@hospitalitaliano.org.ar)}
\thanks{The authors gratefully acknowledge NVIDIA Corporation with the donation of the GPUs used for this research, and the support of UNL (CAID-PIC50420150100098LI, CAID-PIC-50220140100084LI) and ANPCyT (PICT 2016-0651, PICT 2018-03907).}}

\maketitle

\begin{abstract}
Anatomical segmentation is a fundamental task in medical image computing, generally tackled with fully convolutional neural networks which produce dense segmentation masks. These models are often trained with loss functions such as cross-entropy or Dice, which assume pixels to be independent of each other, thus ignoring topological errors and anatomical inconsistencies.  We address this limitation by moving from pixel-level to graph representations, which allow to naturally incorporate anatomical constraints by construction. To this end, we introduce HybridGNet, an encoder-decoder neural architecture that leverages standard convolutions for image feature encoding and graph convolutional neural networks (GCNNs) to decode plausible representations of anatomical structures. We also propose a novel image-to-graph skip connection layer which allows localized features to flow from standard convolutional blocks to GCNN blocks, and show that it improves segmentation accuracy. The proposed architecture is extensively evaluated in a variety of domain shift and image occlusion scenarios, and audited considering different types of demographic domain shift. Our comprehensive experimental setup compares HybridGNet with other landmark and pixel-based models for anatomical segmentation in chest x-ray images, and shows that it produces anatomically plausible results in challenging scenarios where other models tend to fail. 

\end{abstract}


\section{Introduction}

\IEEEPARstart{D}{eep} convolutional neural networks (CNNs) have achieved outstanding performance in anatomical segmentation of biomedical images. Classical approaches employ standard encoder-decoder CNN architectures \cite{ronneberger2015u} that predict the desired segmentation at pixel-level by learning hierarchical features from annotated datasets. Casting image segmentation as a pixel labeling problem is desirable in scenarios where topology and location do not tend to be preserved across individuals, like lesion segmentation. However, organs and anatomical structures usually present a characteristic topology that tends to be regular. Since deep segmentation networks are typically trained to minimize pixel-level loss functions, such as cross-entropy or soft Dice \cite{milletari2016v}, their predictions are not guaranteed to reflect anatomical plausibility, due to the inherent lack of sensitivity that these metrics have with respect to global shape and topology \cite{imagemetrics} (i.e. many different shapes can lead to the same score). Artifacts such as fragmented structures, topological inconsistencies and islands of pixels \cite{bohlender2021survey} are common for such models, especially when faced with challenging real-world clinical scenarios like image occlusions and inter-center domain shift. Incorporating prior knowledge and shape constraints \cite{el2021high} to avoid these artifacts becomes fundamentally important when considering the downstream tasks where segmentation masks are used, like disease diagnosis, therapy planning and patient follow-up. 

As an alternative to dense pixel-level masks, anatomical segmentation can be tackled using other approaches like statistical shape models \cite{heimann2009statistical} or graph-based representations \cite{boussaid2014discriminative}, which provide a natural way to incorporate topological constraints by construction. Such representations make it easier to establish landmark correspondences among individuals, especially important in the context of statistical shape analysis. In particular, graphs appear as a natural way to represent landmarks, contours, and surfaces. By defining the landmark position as a function on the graph nodes, and encoding the anatomical structure through its adjacency matrix, we can easily constrain the space of solutions and encourage topological correctness. 
With the emergence of geometric deep learning \cite{bronstein2017geometric}, CNN extensions to non-euclidean domains like spectral graph convolutions \cite{bruna2013spectral, defferrard2016convolutional} and neural message passing \cite{gilmer2017neural} enabled the construction of deep learning models on graphs. This allowed for the creation of discriminative models that can make predictions based on graph data, as well as deep generative models \cite{kipf2016variational, ranjan2018generating}, which can be used to produce realistic graph structures under a certain distribution.

 \subsubsection*{Contributions} In this work, we explore how landmark-based segmentation can be modeled by combining standard convolutions to encode image features, with generative models based on graph neural networks (GCNNs) to decode anatomically plausible representations of segmented structures. We introduce the HybridGNet architecture, which takes images as input, process them with standard convolutions and then generates landmark-based segmentations by sampling the bottleneck latent distribution, re-shaping and convolving in the graphs domain.
We also present the ``image-to-graph skip connections'' (IGSC) module, which follows the same spirit of UNet skip connections, where feature maps at equivalent resolutions flow from encoder to decoder by-passing the model bottleneck. We propose to extract feature map patches from the image encoder path, and concatenate them with the node features in the GCNN graph decoder to improve accuracy by recovering details from localized high-resolution representations.

A preliminary version of this work was presented at MICCAI 2021\cite{gaggion2021}. In this journal extension we include: 1) The novel IGSC module, which combined with graph unpooling operations, allows localized features at equivalent image/graph resolutions to flow from standard convolutional blocks to GCNN blocks. This is accompanied with new experiments which demonstrate the improvements produced by IGSC when compared to the vanilla HybridGNet and other baselines; 2) A new domain shift study based on two additional datasets, namely the Montgomery and Shenzhen datasets, showing the robustness of HybridGNet to multi-centric domain shift; 3) A new publicly available dataset of landmark annotations with node correspondences generated with HybridGNet for the databases which did not include this type of annotation originally; 4) A new robustness study which includes real X-ray image occlusion from the Padchest dataset caused by pacemakers; 5) A new experimental study to assess the impact of demographic domain shift (in particular, the train/test age distribution) in model performance; 6) Additional experiments to assess the behaviour of the model for pathological anatomy; and 7) A new clinical use-case study where we show how HybridGNet can be used to compute clinically meaningful indices like the cardiothoracic ratio.

\section{Related work}
\begin{figure*}[t]
    \centering
    \includegraphics[width=\linewidth]{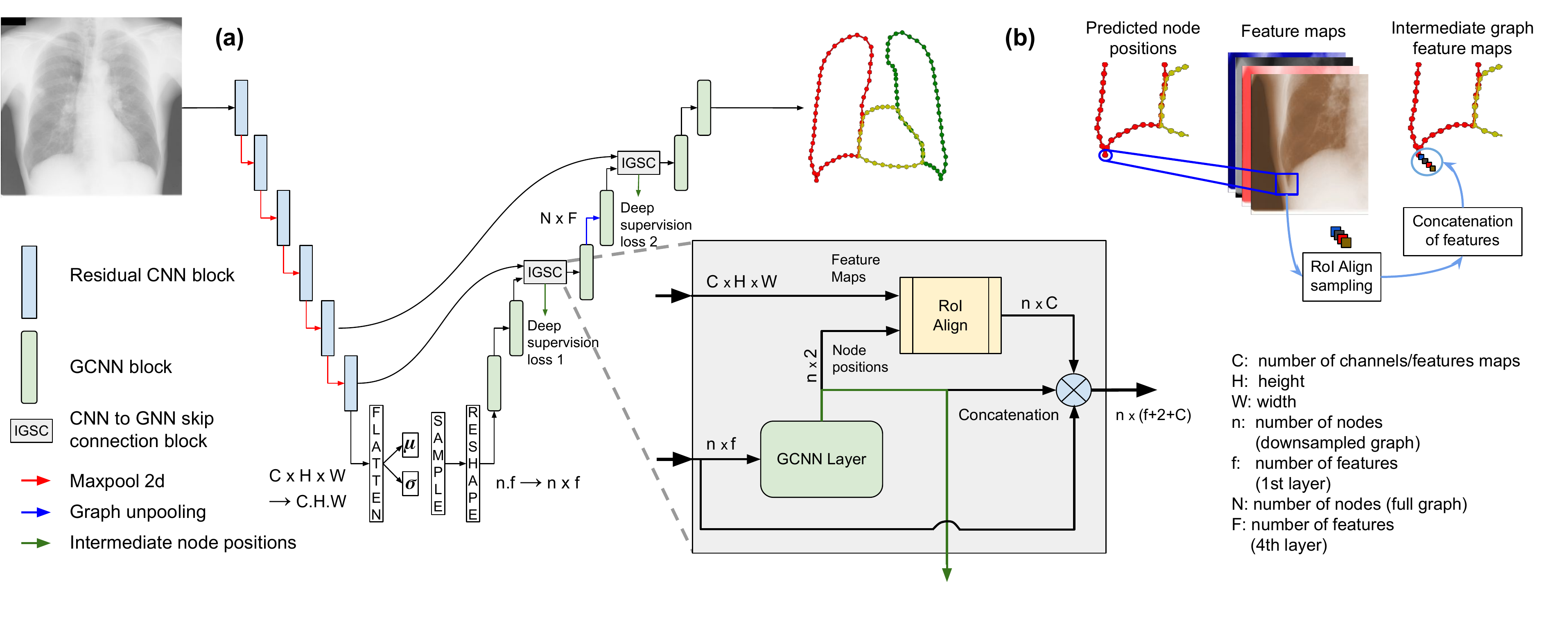}
    \caption{\noindent \textbf{HybridGNet architecture.}
    (a) The proposed HybridGNet architecture combines standard convolutions for image feature encoding (cyan) with graph spectral convolutions (green) to decode plausible anatomical graph-based representations. After the input image is processed by the image encoder, we sample a 1 dimensional vector from the VAE latent distribution (with $n.f$ components) which is then reshaped into a 2D matrix of size $n \times f$ representing the initial node features. Thus, the 1D sampled latent code must have $n.f$ components, so that it can be reshaped into a 2D node features table of dimensions $n \times f$. The Image-to-Graph skip-connection (IGSC) module provides localized features to the intermediate graph representations. (b) Illustrative visualization of the RoIAlign sampling and concatenation of features inside the IGSC module.}
    \label{fig:architecture}
\end{figure*}

\subsubsection*{Landmark-based segmentation} Since the early 1990's, variations of point distribution models (PDMs) have been proposed \cite{cootes1992training} to segment anatomical structures using landmarks. PDMs are flexible shape templates describing how the relative location of important points can vary. Techniques based on PDMs, like active shape models (ASM) \cite{cootes1992training,sozou1997non} and active appearance models (AAM) \cite{cootes1998active} became the defacto standard to deal with anatomical segmentation at the end of the century. Subsequently, the development of more powerful and robust image registration algorithms \cite{zitova2003image} positioned deformable template matching algorithms as the choice of option for anatomical segmentation and atlas construction \cite{frangi2001automatic,heitz2004automatic,paulsen2002building}. In this scenario, contours (for 2D images) and meshes (for 3D images) have been used as deformable templates to solve landmark-based segmentation. However, these methods do not leverage the power of deep neural networks which have dominated image segmentation during the last decade.

More recently, with the advent of deep fully convolutional networks \cite{shakeri2016sub,ronneberger2015u}, major efforts were made to incorporate anatomical constraints into such models \cite{jurdia2020high,larrazabal2019anatomical,oktay2017anatomically}. The richness and robustness of the hierarchical features learned by CNNs allowed them to achieve highly accurate results. Unfortunately, most of these methods work directly on the pixel space, producing acceptable dense segmentation masks, but without landmark annotations and connectivity structure. On the contrary, structured models like graphs can easily represent landmarks, contours and surfaces. In line with this idea, recent studies \cite{milletari2017integrating,bhalodia2018deepssm,bhalodia2021deepssm} have integrated standard CNNs with different representations of landmark structures. These methods employ low-dimensional shape representations like Principal Component Analysis (PCA) decomposition of the original shape space \cite{milletari2017integrating,bhalodia2018deepssm} or performed on more sophisticated particle distribution models \cite{bhalodia2021deepssm}. In this work, inspired by previous studies on graph generative models \cite{ranjan2018generating}, we propose to replace such embeddings by more powerful non-linear representations based on hierarchical graph convolutional \cite{bronstein2017geometric} decoders. 

\subsubsection*{Graph generative models} We want to exploit the generative power of graph variational autoencoders \cite{foti2020intraoperative} to decode plausible anatomical segmentations from low dimensional embeddings. Of particular interest for our work is the convolutional mesh autoencoder proposed in \cite{ranjan2018generating}. The authors constructed an encoder-decoder network using spectral graph convolutions, and trained it in a variational setting using face meshes. By sampling the latent space, they are able to generate new expressive faces, never seen during training. We build on top of this idea by keeping the graph convolutional \textit{decoder}, but replacing the graph \textit{encoder} with a standard CNN-based encoder that takes images as inputs. This hybrid architecture learns a variational distribution conditioned on image data, from which we can sample graphs representing anatomically plausible segmentations.

\subsubsection*{Image-to-graph localized skip connections} Last but not least, we are interested in producing accurate landmark-based segmentation for high-resolution 2D images. In that sense, propagating features learned at different hierarchical levels from encoder to decoder through skip connections has shown to be an effective mechanism not only to improve segmentation accuracy, but also to increase convergence speed and enable training of very deep networks \cite{drozdzal2016importance}. Previous approaches incorporated different types of skip-connections in the context of mesh extraction from images. Pixel2Mesh \cite{pixel2mesh} introduces a perceptual feature pooling layer designed to work with 3D meshes and 2D images, thus projecting 3D vertices to the image plane using camera intrinsics, which does not apply for our case where input image and output graph live in the same 2D space. Closer to our approach is Voxel2Mesh \cite{wickramasinghe2020voxel2mesh}, a model designed to operate on images and graphs living in the same dimension. Voxel2Mesh employs a learned neighborhood sampling layer which pools image features in locations indicated by the node coordinates. However, both Voxel2Mesh and Pixel2Mesh build on the idea of deforming an initial sphere mesh template, thus limiting its applicability to certain topologies and single object segmentation. Another approach based on template refinement is Curve-GCN\cite{ling2019fast}, an algorithm that can be used to perform automatic or interactive 2D landmark-based segmentation, which relies on the deformation of an initial ellipsoid template contour. 
Curve-GCN operates based on an iterative approach where the contour node displacements are successively inferred and used to deform an initial template. In this case, only localized image features are sampled from the corresponding node positions.
On the contrary, our model employs a generative approach which samples a global embedding from a variational latent distribution, which is then used to directly decode the node positions in a single forward pass.
Other approaches resort to refining meshes obtained from voxel predictions \cite{meshRCNN}. Here we adopt a different approach where output graphs (2D contours in our case) do not correspond to a deformed template, but instead are directly sampled from a latent distribution learnt during training. We also propose a new image-to-graph skip connection (IGSC) layer based on the well-known RoIAlign module \cite{maskRCNN}, which enables end-to-end learning of localized features guided by intermediate node coordinates.

\section{Anatomical segmentation via hybrid graph neural networks}

\subsection{Preliminaries}
\label{sec:preliminaries}

\subsubsection*{Problem setting} Let us have a dataset $\mathcal{D} = \{(\mathbf{I}, \mathbf{G})_{n}\}_{0 < n \leq N}$, composed of $N$ images $\mathbf{I}$ and their corresponding landmark-based segmentation represented as a graph $\mathbf{G} = \langle V,\mathbf{A},\mathbf{X} \rangle$ (see Section \ref{sec:graphconstruction} for a detailed description of the graph construction for the heart and lungs structures used in this study). $V$ is the set of nodes for $M$ landmarks, $\mathbf{A} \in \{0,1\}^{M \times M}$ is the adjacency matrix indicating the connectivity between pairs of nodes ($a_{ij} = 1$ indicates an edge connecting vertices $i$ and $j$, and $a_{ij} = 0$ otherwise) and $\mathbf{X} \in \mathbb{R}^{M \times s}$ is a function (represented as a matrix) assigning a feature vector to every node. In our case, it assigns a 2-dimensional spatial coordinate (the landmark position) to every node ($s=2$). In the context of landmark-based segmentation and point distribution models, it is common (and useful) to have manual annotations with a fixed number of points. Therefore, we assume that $V$ and $\mathbf{A}$ are the same for all the images in the dataset, the only difference among them is given by the spatial coordinates defined in $\mathbf{X}$. This assumption enables us to follow the work of \cite{defferrard2016convolutional,ranjan2018generating} and use spectral graph convolutions to learn latent representations of anatomy. Please note that, following the literature of landmark-based segmentation in medical imaging \cite{gaggion2021,besbes2011landmark}, we use the term \textit{landmark} to denote points that can be uniquely identified in a set of shapes. However, these are also called \textit{pseudo-landmarks} as not all of them are actual anatomical points, but they are landmarks lying on the contour of a shape, determining its geometry \cite{frangi2001automatic}.

\subsubsection*{Spectral graph convolutions} 
Spectral convolutions are built using the eigendecomposition of the graph Laplacian matrix $\mathbf{L}$, exploiting the property that convolutions in the node domain are equivalent to multiplications in the graph spectral domain \cite{shuman2013emerging}. 
The graph Laplacian is defined as $\mathbf{L} = \mathbf{D} - \mathbf{A}$, where $\mathbf{D}$ is the diagonal degree matrix with $d_{ii} = \sum_{j} a_{ij}$.
The Laplacian can be decomposed as $\mathbf{L} = \mathbf{U \Lambda U}^T$, where $\mathbf{U} \in \mathbb{R}^{M \times M} = [\mathbf{u}_0, \mathbf{u}_1, \ldots, \mathbf{u}_{M-1}]$ is the matrix of eigenvectors (Fourier basis) and $\mathbf{\Lambda} = \mathrm{diag}(\lambda_0, \lambda_1, \ldots, \lambda_{M-1})$ is the matrix of eigenvalues (frequencies of the graph).
By analogy with the classical Fourier transform for continuous or discrete signals, the graph Fourier transform of a function $\mathbf{X}$ defined on the graph domain can be obtained as $\mathbf{\hat{X}} = \mathbf{U}^T \mathbf{X}$, while its inverse is given by $\mathbf{X} = \mathbf{U} \mathbf{\hat{X}}$.
Based on this formulation, the spectral convolution between a signal $\mathbf{X}$ and a filter $\mathbf{g}_\phi = \mathrm{diag}(\boldsymbol{\phi})$ is defined as $\mathbf{g}_\phi * \mathbf{X} = \mathbf{g}_\phi \mathbf{L} \mathbf{X} = \mathbf{g}_\phi (\mathbf{U \Lambda U}^T) \mathbf{X} = \mathbf{U} \mathbf{g}_\phi \mathbf{\Lambda} \mathbf{U}^T \mathbf{X}$, where $\boldsymbol{\phi} \in \mathbb{R}^n$ is a vector of coefficients parameterizing the filter.
We follow the work of Defferrard et al \cite{defferrard2016convolutional} and restrict the class of filters to polynomial filters with the form $\mathbf{g}_\phi = \sum_{k=0}^K \phi_k \mathbf{\Lambda}^k$. Polynomial filters are strictly localized in the vertex domain (a K-order polynomial filter considers K-hop neighborhoods around the node) and reduce the computational complexity of the convolutional operator. Such filters can be well approximated by a truncated expansion in terms of Chebyshev polynomials, computed recursively. Following  \cite{defferrard2016convolutional,ranjan2018generating} we adopt this approximation to implement the spectral convolutions. Note that a spectral convolutional layer will take feature matrices $\mathbf{X}^\ell$ as input and produce filtered versions $\mathbf{X}^{\ell+1}$, similar to what standard convolutions do with images and feature maps.

\subsection{HybridGNet: Image-to-graph extraction via hybrid convolutions}

The proposed neural network takes images as input and produces graphs as output, combining standard with spectral convolutions in a single model that is trained end-to-end. The current HybridGNet formulation follows the same principles introduced in our original MICCAI publication \cite{gaggion2021}, but incorporates new elements like image-to-graph skip connections, graph unpooling operations and variations in the training strategy, that will be later highlighted. Let us start by defining the basic architecture, which resembles a variational autoencoder (VAE) \cite{kingma2013auto} (see Figure \ref{fig:architecture}) in the sense that the latent space models a variational distribution parameterized as a multivariate Gaussian. 

Autoencoders are neural networks designed to reconstruct their input. They follow an encoder-decoder scheme, where an encoder $\mathbf{z} = f_{e}(\mathbf{I})$ maps the input image $\mathbf{I}$ to a lower dimensional latent code $\mathbf{z}$, which is then processed by a decoder $f_{d}(\mathbf{z})$ to reconstruct the original input. 
The bottleneck imposed by the low-dimensionality of the encoding $\mathbf{z}$ forces the model to retain useful information, learning powerful representations of the data distribution. The model is trained to minimize a reconstruction loss $\mathcal{L}_{r}(\mathbf{I}, f_{d}(f_{e}(\mathbf{I})))$ between the input and the output reconstruction. To constrain the distribution of the latent space $\mathbf{z}$, we add a variational term to the loss function, resulting in a variational autoencoder (VAE) \cite{kingma2013auto}.  
We assume that the latent codes $\mathbf{z}$ are sampled from a distribution $Q(\mathbf{z})$ for which we will impose a unit multivariate Gaussian prior. In practise, during training, this results in the latent codes $\mathbf{z}$ being sampled from a distribution $\mathcal{N}(\mathbf{\mu}, \mathbf{\sigma})$ via the reparametrization trick \cite{kingma2013auto}, where $\mathbf{\mu}$, $\mathbf{\sigma}$ are deterministic parameters generated by the encoder $f_{e}(\mathbf{I})$. Given a sample $\mathbf{z}$, we can generate (reconstruct) the corresponding data point by using the decoder $f_{d}(\mathbf{I})$. This model is usually trained by minimizing a loss function defined as:
\begin{equation}
    \mathcal{L}_{a} = \mathcal{L}_{r}(\mathbf{I}, f_{d}(\mathbf{z})) + w\  \textrm{KL}\left(\mathcal{N}(0,1) || Q(\mathbf{z} | \mathbf{I})\right),
\end{equation}
where the first term is the reconstruction loss, the second term imposes a unit Gaussian prior $\mathcal{N}(0,1)$ via the KL divergence loss and $w$ is a weighting factor. 

In our previous work \cite{gaggion2021}, HybridGNet was constructed by first pre-training two independent VAEs with the same latent dimension: one to reconstruct images using standard convolutions $f^\mathcal{I}(\mathbf{I}) = f^\mathcal{I}_{d}(f^\mathcal{I}_{e}(\mathbf{I}))$ and another one to reconstruct graphs via spectral convolutions $f^\mathcal{G}(\mathbf{G}) = f^\mathcal{G}_{d}(f^\mathcal{G}_{e}(\mathbf{G}))$. Once both models were trained, we decoupled their encoders and decoders, keeping only the image encoder $f^\mathcal{I}_{e}(\mathbf{I})$ and graph decoder $f^\mathcal{G}_{d}(\mathbf{z})$. The HybridGNet was then constructed by connecting these two pre-trained networks as $f^{\mathcal{H}}(\mathbf{I}) = f^\mathcal{G}_{d}(f^I_{e}(\mathbf{I}))$ and re-training until convergence by minimizing:
\begin{equation}
    \mathcal{L}_{\mathcal{H}} = \mathcal{L}_{r}(\mathbf{G}, f^\mathcal{G}_{d}(\mathbf{z})) + w\  \textrm{KL}\left(\mathcal{N}(0,1) || Q(\mathbf{z} | \mathbf{I})\right),
\end{equation}

\noindent where $\mathcal{L}_{r}(\mathbf{G}, f^\mathcal{G}_{d}(\mathbf{z}))$ is the graph-reconstruction loss computed as the mean squared error (MSE) of the predicted node positions, and $Q$ is the variational distribution parameterized by $f_e^\mathcal{I}(\mathbf{I})$.

Here we simplify the training strategy by eliminating the pre-training stage and directly training $f^{\mathcal{H}}(\mathbf{I})$ from scratch, since we observed that pre-training only helps to achieve faster convergence, but does not produce significant improvements in terms of segmentation accuracy. This simplified end-to-end training process directly learns a single latent space relating images and graphs.

\subsubsection*{Graph unpooling}
We included a fixed graph unpooling layer in the graph decoder $f^\mathcal{G}_{d}(\mathbf{z})$, to learn representations at multiple resolutions \cite{ranjan2018generating}. We adopted a simple strategy where all graphs $\mathbf{G}$ in our dataset are pre-processed to produce lower resolution graphs $\mathbf{G}^k$ by reducing to half the number of nodes $k$ times, replacing pairs of consecutive neighboring nodes with a single one, whose position is computed as their average. The unpooling layer is defined so that it reverses this operation by duplicating the number of nodes and interpolating the features between them. The unpooling layer was included after the 3rd GCNN layer of the decoder as shown in Figure \ref{fig:architecture}.

\subsubsection*{Localized image-to-graph skip connections (IGSC) and deep supervision}

Under the hypothesis that local image features may help to produce more accurate estimates of landmark positions, we designed a localized Image-to-Graph Skip Connection (IGSC) layer (see Figure \ref{fig:architecture}.a,b). IGSC uses the well-known RoIAlign module \cite{maskRCNN} to sample localized features for each node from a specific encoder level given a certain location, which in our case is specified by intermediate node positions learned via deep supervision \cite{deepsupervision}. This layer is parameterized by a window size, indicating the area that will be sampled for every node. It receives a tensor of feature maps and a list of node positions which indicate the spatial location from where the feature map will be sampled, and returns the corresponding regions of interest (RoIs) of the given window size centered at the node positions. In our model, an internal GCNN layer learns intermediate node positions via deep-supervision, resulting in extra loss terms $\mathcal{L}_{DS}$ which compute the mean squared error between the ground truth node position (for both graph resolutions) and the intermediate predictions. The desired window input size was set to 3x3, while the output size was set to 1x1, so it only returns a single value per feature-map, which is calculated using average pooling. Then, this array of features is concatenated with the original node features and an augmented graph is obtained as shown in Figure \ref{fig:architecture}.b).

\section{Experimental setup}

\subsection{Database description}
\label{sec:databases}
We evaluated the proposed model in a variety of tasks involving chest x-ray image segmentation. In what follows, we describe the databases used to perform these experiments.

\subsubsection{JSRT Database}
\label{sec:graphconstruction}

The Japanese Society of Radiological Technology (JSRT) Database \cite{shiraishi2000development} consists on 247 high resolution x-ray images, with expert landmark annotations (120 landmarks per image) for lung and heart \cite{gt_van_ginneken:2006-1223}. The image resolution was 1024x1024 px, with a pixel spacing of 0.35x0.35 mm. The dataset was randomly split into 70\%-10\%-20\% partitions for training, validation and test, respectively. 

Given the mask contour for the structures of interest, several anatomical landmarks in the lung and heart were manually identified (e.g. the lung apex): 4 for the right lung, 5 for the left lung, and 4 for the heart \cite{gt_van_ginneken:2006-1223}. The other intermediate points were interpolated across the mask contour providing a final set of 44, 50 and 26 points respectively. The number of interpolation points between anatomical landmarks was fixed for all subjects, resulting in a one-to-one correspondence of the distinctive points between subjects.
The graphs described in Section \ref{sec:preliminaries} were then constructed by taking each landmark as a node, defining edges between neighboring nodes, and the $(x,y)$ positions of each landmark were set as the node features. The adjacency matrix was then pre-computed to represent the connectivity structure of the aforementioned graph. By having the same number of landmarks with one-to-one correspondences, the adjacency matrix was the same for all graphs.

\subsubsection{Montgomery County and Shenzhen Hospital x-ray sets}
Two public chest x-ray datasets with dense lung segmentation masks were used as external test sets to evaluate inter-dataset DS. The Montgomery County dataset (138 images) \cite{montgomeryset} was acquired from the tuberculosis control program of the Department of Health and Human Services of Montgomery County, MD, USA. The Shenzhen dataset (566 images) \cite{shenzhen} was collected as part of the routine care at Shenzhen No.3 Hospital in Shenzhen, Guangdong providence, China. 

\subsubsection{Padchest dataset}
Consists of 160,868 chest x-ray images from 67,000 patients \cite{bustos2020padchest} including labels for 174 radiological findings, 19 diagnostic labels, and 104 anatomic locations. Although this dataset does not contain segmentation masks, a subset of 137 images with \emph{cardiomegaly} diagnosis label were manually segmented by two radiologists who delineated the lungs and heart as dense masks, to evaluate our method in a real clinical task, namely cardiothoracic ratio estimation. 
From these images, 20 included pacemakers and 45 also included an \emph{aortic elongation} label. The images with pacemakers were used to evaluate the robustness of the proposed model to occlusions produced by external artifacts.

\input{table_JSRT.tex}

\begin{figure*}[th] 
    \centering
    \includegraphics[width=0.95\linewidth]{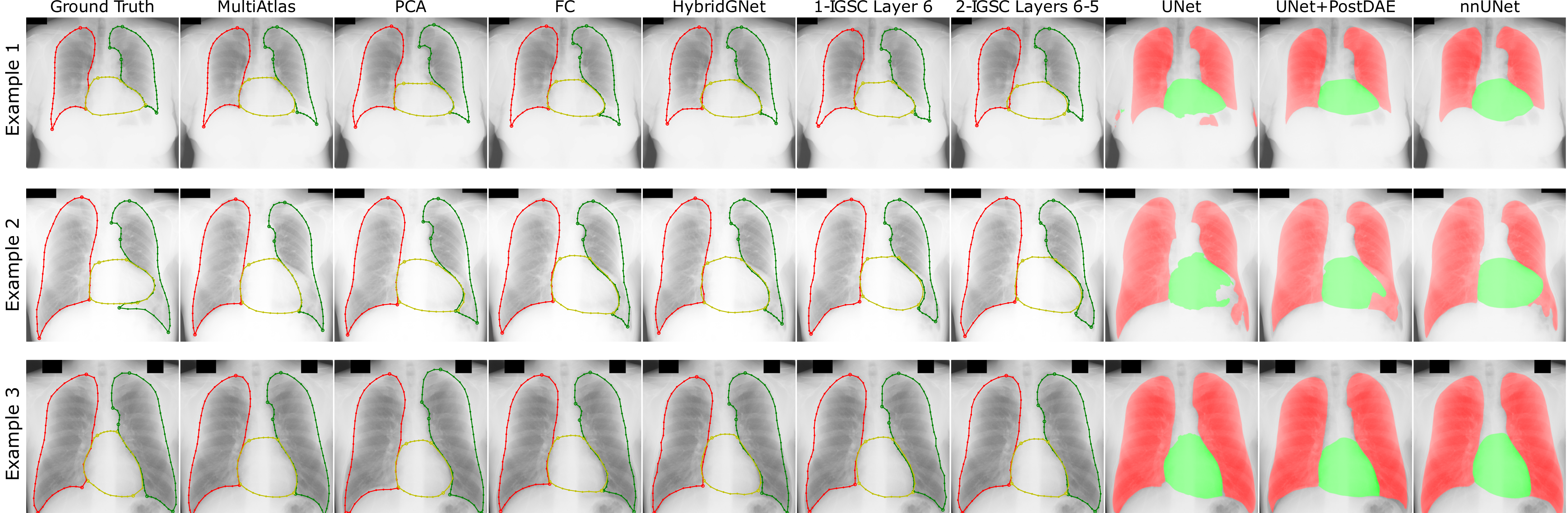}
    \caption{\noindent \textbf{Landmark-based anatomical segmentation.}
    Qualitative analysis for the JSRT test set. Results reflect the improvement in anatomically plausibility obtained when using the HybridGNet with IGSC.}
    \label{fig:compare_jsrt}
\end{figure*}

\subsection{Baselines models}

Our work builds on the hypothesis that encoding connectivity information through graph structures can provide richer representations than standard landmark-based point distribution models. To evaluate this hypothesis, we build standard point distribution models from the graph representations by considering landmarks as independent points. For a given graph $\mathbf{G} = \langle V, \mathbf{A}, \mathbf{X} \rangle$, we construct a vectorized representation by concatenating the rows of $\mathbf{X}$ in a single vector as 
$\boldsymbol{\rho} = \left[x_{0,0}, x_{0,1}, x_{1,0}, x_{1,1}, \ldots, x_{M-1,0}, x_{M-1,1}\right]$.

\subsubsection{PCA} We first consider a single baseline similar to \cite{milletari2017integrating,bhalodia2018deepssm}, by performing principal component analysis (PCA) to transform the vectorized representation $\boldsymbol{\rho}$ into lower-dimensional embeddings. We then optimize the CNN encoder $f^\mathcal{I}_{e}$ to estimate the PCA coefficients, reconstructing the landmarks as a linear combination of the principal components.
\subsubsection{FC}The second baseline combines the CNN encoder $f^\mathcal{I}_{e}$ with a fully connected (FC) decoder that directly reconstructs the vectored representations $\boldsymbol{\rho}$. 

\subsubsection{Multi-atlas}The third baseline implements a multi-atlas segmentation approach \cite{alven2019shape,alven2016shape}, which employ several labeled atlases (i.e. pairs consisting of an image and its associated landmark-based segmentation) to delineate the structures of interest. Given a target image to be segmented, the 5 atlases most similar to the target image (based on the mutual information metric) are obtained from the training set. Then, we perform pairwise non-rigid registration (with affine initialization) using SimpleElastix \cite{marstal2016simpleelastix}. Registration allows to transfer the landmarks of each selected image into the target space. The final landmark-based segmentation is obtained by averaging the position of the set of candidate landmarks.

\subsubsection{UNet} Finally, a UNet \cite{ronneberger2015u} model was also included to benchmark our approach against a standard pixel-level segmentation method. We used the CNN encoder $f^\mathcal{I}_{e}$ and decoder $f^\mathcal{I}_{d}$ with standard skip-connections via concatenation, to guarantee comparable complexity. 

\subsubsection{Post-DAE Postprocessing:} UNet results were post-processed using a denoising autoencoder following \cite{larrazabal2019anatomical}, which was trained to obtain plausible segmentations from noisy dense segmentation masks.

\subsubsection{nnUNet} a state-of the art nnUNet \cite{isensee2021nnunet} was included, which consists of a self-configuring pipeline of automatic pre-processing, hyperparameter search, training, ensembling and post-processing following a 5-fold cross validation scheme. This 5-fold set is then treated as a ensemble model, averaging the predictions and incorporating standard post-processing techniques.

\subsection{Implementation and training details}
All models were implemented in PyTorch\cite{pytorch}, using PyTorch Geometric \cite{pytorch-geometric} for the spectral GCNN layers\footnote{Source code is publicly available at \url{https://github.com/ngaggion/HybridGNet}. The experiments are saved on Jupyter notebooks and extra information on statistical significance is also available in the repository. The Multi-atlas implementation is available at \url{https://github.com/lucasmansilla/multiatlas-landmark}.}.
Every model and baseline shares the same CNN encoder $f^\mathcal{I}_{e}$, with 6 residual blocks \cite{resnet-paper} interleaved with max-poolings as shown in Figure \ref{fig:architecture}. For the GCNN decoders, we use 6 layers of Chebyshev convolutions with Layer Normalization \cite{ba2016layer} and ReLU nonlinearities. We set the k-hop neighbourhood parameter for the graph convolutions at 6. This hyperparameter was chosen performing an ablation study on the validation data, which is not included for space restrictions, but it is available in our repository. For HybridGNet models, we evaluated the inclusion of 1 and 2 IGSC modules, extracting features from layers 3 to 6 of the encoder. We also evaluated the incorporation of a third IGSC module (consequently adding another graph downsampling level), but it resulted in slightly worse performance (due to the low resolution of the downsampled graphs). Thus, we decided to stick to 2 IGSC modules.

\subsubsection{Data augmentation}
Online data augmentation was used to train all the models (i.e. baselines and HybridGNet) including: 
i) bright augmentation using a Gamma correction with random gamma between 0.60 and 1.40;
ii) random image rotations between -3 and 3 degrees;
iii) vertical and horizontal random scaling, ensuring that landmarks remain inside the visible area;
iv) cropping or padding the images randomly if the shape was different to the expected input shape ($1024 \times 1024$).

\subsubsection{Model training} All models were trained for 3000 epochs using Adam optimizer, with a learning rate of 1e-4, a batch size of 4, a weight decay of 1e-5, and a KL divergence weight factor $w = $1e-5. To prevent overfitting, learning rate decay was set to reduce it by 0.9 every 100 epochs (for IGSC models) and by 0.9 every 50 epochs (for HybridGNet model). We used the MSE in pixel space over the vectored landmark location as loss function for landmark models, and a combination of Dice and cross-entropy for the UNet. Checkpoints were selected based on validation loss.

\section{Experiments and discussion}
We performed a series of experiments to compare the proposed HybridGNet and its variants with the aforementioned baselines, and evaluate their performance in a variety of scenarios and tasks.\\

\subsubsection{Model comparison}
\label{IGSC-layer-ablation}First, we compared HybridGNet with the baselines using the JSRT dataset, and assessed the effect of skip connections by evaluating alternative HybridGNet architectures. 
We used metrics that can be derived from graph representations, including landmark MSE and Hausdorff distance (HD, in millimeters). To benchmark our methods against pixel-level methods, we filled the organ contours to obtain dense masks from graph representations, and computed the Dice coefficient.

Table \ref{results-jsrt} reports metrics on the test set of JSRT dataset (bold numbers indicate the best performance for methods of the same type, i.e. landmark and pixel-level methods). All differences are significant according to the Wilcoxon test (except for the 1-IGSC Layer 6 model that has no significant difference in HD Heart with the 2-IGSC L6-5 model). First, it is worth noting that when comparing HybridGNet models with and without skip connections, there is a big difference in terms of MSE and HD in favor of the model with 2 IGSC (Layers 6-5), implying that localized features help to improve landmark prediction accuracy. Moreover, the HybridGNet 2 IGSC (Layers 6-5) outperforms the landmark-based baselines on MSE, Dice, and HD, confirming our main hypothesis that incorporating graph connectivity structure helps in producing more realistic segmentations.

For the sake of completeness, we also included pixel-level segmentation baselines. HybridGNet surpasses UNet and UNet+PostDAE baselines by a large margin in terms of HD and it is competitive with the nnUNet in that regard. On the contrary, the UNet model and the self-configuring nnUNet outperform the HybridGNet variants in Dice, what is somehow expected since dense predictions are not directly optimized in our models. In that vein, while Dice is agnostic to topological errors and islands of pixels (in the sense that wrong predictions are penalized independently of their location), due to its formulation HD is more sensitive to them, better reflecting anatomical plausibility, which is the main interest of this work (see \cite{imagemetrics} for a complete discussion about the limitations of image segmentation metrics). 
Figure \ref{fig:compare_jsrt} shows qualitative results for 3 exemplar cases and Table \ref{inference_times} shows inference times for HybridGNet and pixel-based models on an NVIDIA Titan Xp with 12GB RAM. The lower inference time of the HybridGNet is mainly due to the fact that the number of parameters in the graph decoder is much lower than that of the standard convolutional decoder.

\input{table_inference_time}

As an important remark, although we include state-of-the-art pixel level segmentation methods like UNet, nnUNnet and Post-DAE for completeness, we highlight the fact that our method is a landmark-based segmentation approach which produces structured predictions in the form of graphs (not pixel-level masks), enabling the extraction of additional information like anatomical landmark locations and inter-patient node correspondences, which cannot be obtained with standard pixel-level segmentation methods. Thus, for a more fair comparison of the HybridGNet results, the reader should focus on the landmark-based MSE which can only be computed for landmark-based segmentation methods like FC, PCA, Multi-atlas and the different variants of the HybridGNet model, which share the same output representation.

\subsubsection{Generating landmark-based representations from dense segmentations}
\label{sec:generatingLandmarks}
In this work we considered landmark-based segmentations with a fixed number of points, that enable establishing correspondences across images. This hinders the applicability of our method to tasks like brain tumor segmentation, where the shape of the structures of interest is not regular. Nonetheless, this is desirable in scenarios like population shape analysis, where we are interested in understanding how certain anatomical keypoints vary for different individuals. Unfortunately, in most segmentation datasets, only pixel-level annotations are available. In these cases, one could ask expert medical doctors to manually annotate specific landmarks in the mask contour (see Section \ref{sec:graphconstruction} as an example) or perform automated estimation of landmarks from dense segmentations using HybridGNet. Our model can be trained to recover landmark-based representations from dense segmentation masks in a natural way. Thus, we trained our best performing models and baselines with dense segmentation masks as input (instead of images), to perform landmark estimation. Table \ref{tables2g} shows the results on the JSRT test set: the proposed HybridGNet 2 IGSC (Layers 6-5) outperforms the other baselines and architectures, proving useful in the building of shape models with landmark correspondences from pixel-level masks. Multi-atlas showed no differences in Dice with respect to our HybridGNet 2 IGSC (according to Wilcoxon's test), but we observed that it loses track of the point-to-point correspondences as it is exhibited by the higher MSE error, which is computed for pairs of matching points. 

HybridGNet was used to create landmark annotations for the Montgomery, Shenzhen and Padchest datasets, which originally did not include this type of segmentations. We are publicly releasing these new annotations\footnote{Annotations available at: \url{https://github.com/ngaggion/Chest-xray-landmark-dataset}} hoping that they will serve for future studies where point correspondences across individuals are required.\\

\input{table_S2G.tex}

\subsubsection{Domain shift (DS) evaluation}

\input{table_Montgomery_Shenzhen.tex}
DS refers to a variation in the target (test) domain concerning the source (training) domain \cite{choudhary2020advancing}. In most cases, such DS drops performance significantly as supervised learning assumes that training samples have the same distribution as the test samples. DS can be caused by multiple factors including changes in acquisition parameters, medical center or population demographics. We compared the effect of DS by measuring segmentation performance on datasets captured at different medical centers, i.e. training in the JSRT dataset and testing with Shenzhen and Montgomery. Table \ref{results-mont-shen} shows how HybridGNet model greatly outperforms most baselines in terms of HD and Dice, and it is competitive with the self-configuring nnUNet (bold numbers indicate the best performance for methods
of the same type, i.e. landmark and pixel-level methods). Differences between the two best performing methods (nnUNet and 2 IGSC: L6-5) and all other baselines are significant according to Wilcoxon test. On the contrary, differences between nnUNet and 2 IGSC: L6-5 are not significant, except for Dice Lung in the Shenzen dataset. These results confirm the generalizability of the proposed model across medical centers.

\begin{figure}
    \centering
    \includegraphics[width=\linewidth]{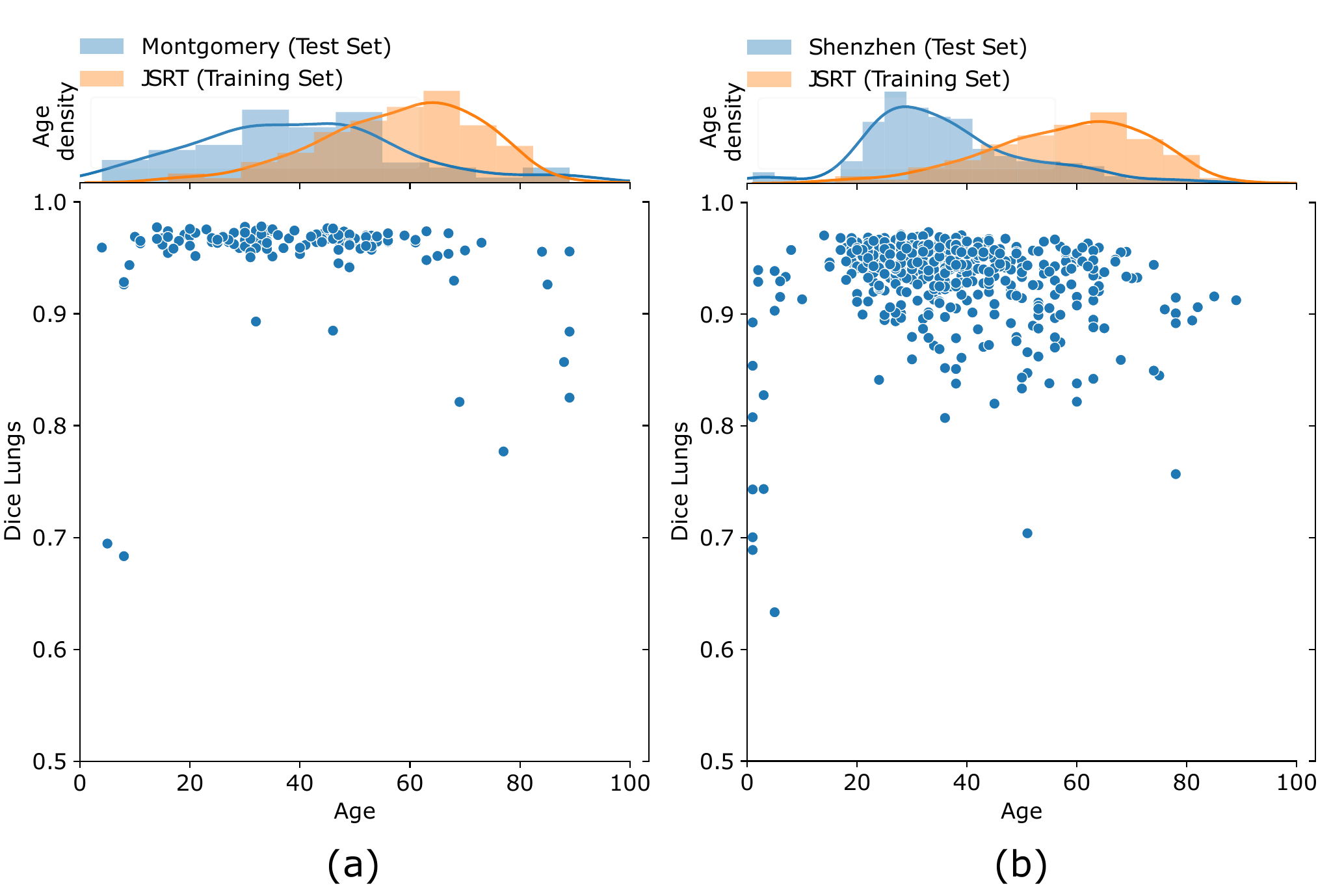}
    \caption{\noindent \textbf{Assessing the impact of domain shift by age distribution on lung segmentation.}
    Scatter plot of the lung Dice coefficient vs. age of patient for every individual in both (a) Montgomery and (b) Shenzhen datasets. Histograms show the age distribution for test (blue) and training sets (orange).}
    \label{fig:ages}
\end{figure}

Moreover, recent studies on fairness in machine learning have shown that under-representation of certain demographic groups in the training data (e.g. in terms of gender \cite{larrazabal2020gender} or ethnicity \cite{puyol2021fairness}) may result in biased models which present unequal performance in minority groups. Here we are interested in evaluating if the same holds for chest x-ray segmentation, in particular when considering age distribution shifts between training and test patients. To perform this analysis, we take our best performing model (HybridGNet 2 IGSC Layers 6-5) and build a scatter plot (see Figure \ref{fig:ages}) depicting the Dice coefficient for lung segmentation vs patients age. When observing the age histograms between training and test sets, we note that young patients are highly underrepresented in the training set. Interestingly, we found that model performance drastically drops for patients between 0-18 years old in both Montgomery and Shenzen datasets, what can be attributed to the lack of young people on the JSRT database. Since the size of the organs tends to be smaller for younger patients (in particular we observed significant differences for the lung area), this can bias the learning process if this subpopulation is not well represented. This experiment highlights the importance of performing disaggregated analysis to detect potential subgroups where the model may under-perform.

\subsubsection{Robustness to image occlusions (IO)}
\label{sec:occlusions}

IO are common in chest x-rays, for example due to patient de-identification, that is covering protected information with black patches, or devices such as pacemakers, tubes, cables, or electrodes that can cover important parts of organs or other structures of interest. This is a common situation found in clinical centers, especially in images from hospitalized patients. We designed two experiments to assess the robustness of HybridGNet to artificial and real IO that were not represented in the training set, and compare it with the other baselines.

\begin{figure}
    \centering
    \includegraphics[width=\linewidth]{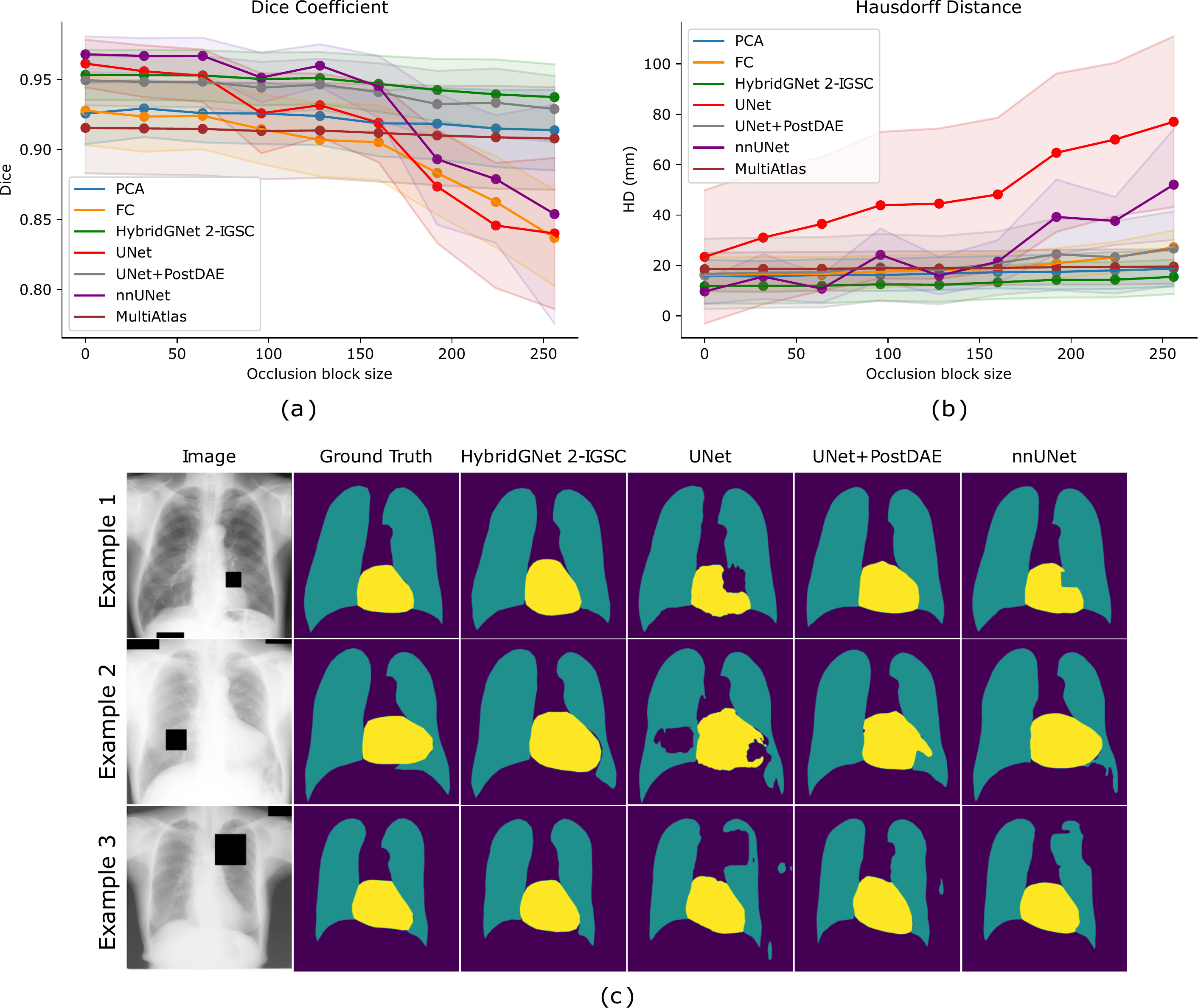}
    \caption{\noindent \textbf{Artificial occlusions study.}
    (a) Dice and (b) HD distance for increasing block size in artificial occlusions.
    (c) Shows qualitative results.}
    \label{fig:artificial}
\end{figure}

First, we simulated artificial occlusions by overlapping a random black box on every image. We applied boxes of different sizes over the JSRT test set on random positions. Figures \ref{fig:artificial} (a) and (b) show Dice and HD distance for lungs and heart segmentation (averaged) as the occlusion block size increases. Although both nnUNet and UNet slightly outperform HybridGNet in Dice for very small oclussions, its performance drops with a steeper slope than HybridGNet as we increase the size of the occlusion block. Figure \ref{fig:artificial} (c) shows some qualitative results for three cases with different occlusion levels. Both quantitative and qualitative results show that HybridGNet is more robust to IO than pixel-level models.

\begin{figure}
    \centering
    \includegraphics[width=\linewidth]{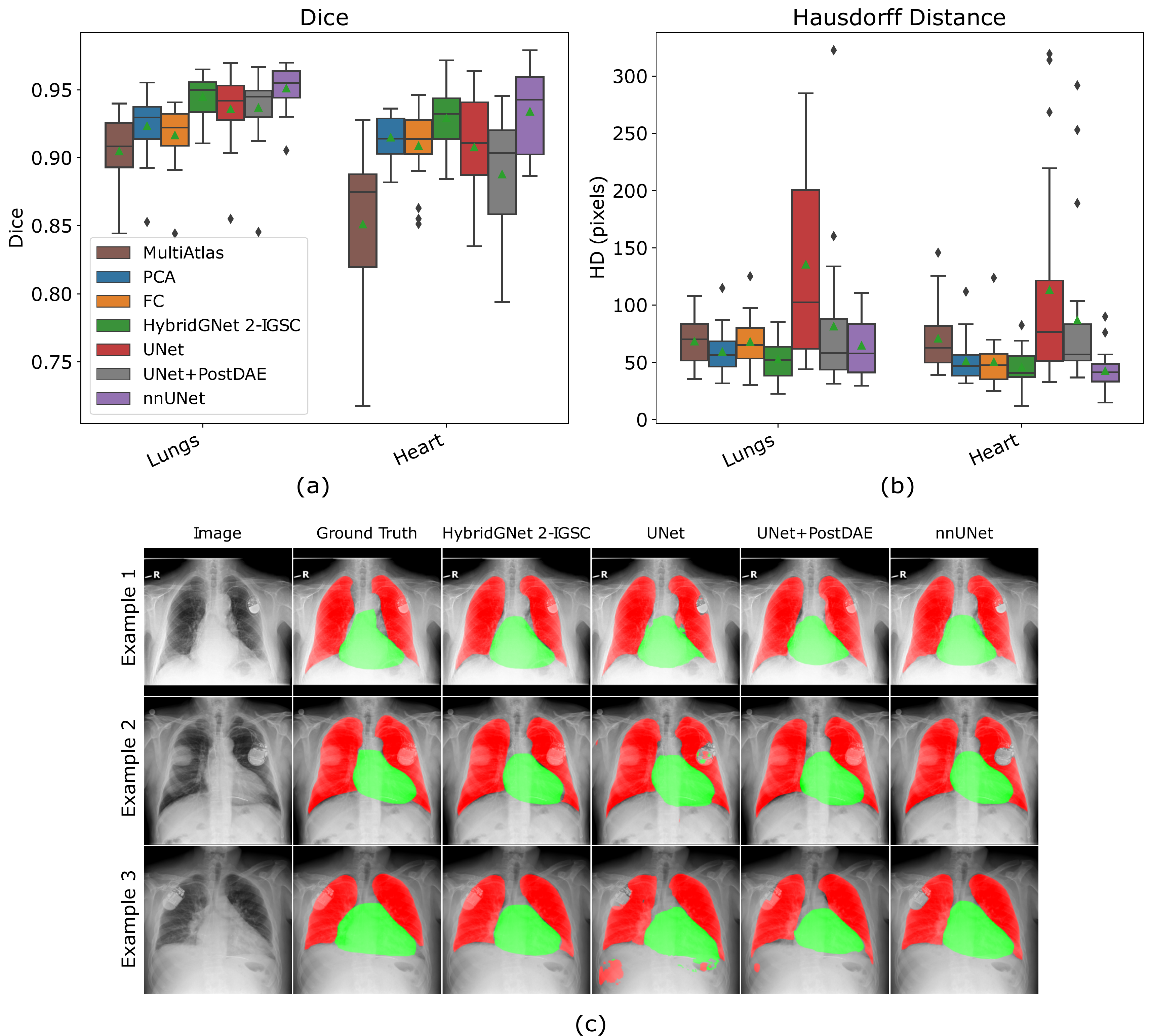}
    \caption{\noindent \textbf{Real occlusions study.}
    (a) Dice and (b) HD distances for the pacemaker Padchest subset. Wilcoxon test showed no significant differences between HybridGNet and nnUNet, the best two performing models. However, qualitative results (c) show how HybridGNet results in more anatomically plausible segmentations than nnUNet for complex cases.
}
    \label{fig:realoc}
\end{figure}

Robustness to real occlusions produced by external devices was also assessed. To this end, we used 20 segmented images with pacemakers from Padchest as test set. To evaluate solely the occlusion effect on performance and alleviate DS issues due to intensity differences across different medical centers, we retrained the models (both HybridGNet and baseline) with an extended training dataset that includes Padchest images (without pacemakers). In Figure \ref{fig:realoc} we can see how our model outperforms the UNet both on Dice and Hausdorff distance, while having no significant differences when compared to the nnUNet but producing more anatomically plausible images qualitatively.\\

\subsubsection{Model behaviour on pathological anatomy.} We are also interested in analyzing the behaviour of HybridGNet in the context of pathological anatomy. To this end, we followed the experimental setup introduced in \cite{larrazabal2019anatomical} where a subset of patients from the Shenzhen database diagnosed with tuberculosis was considered. These patients have a collapsed lung and therefore a reduced air cavity. Every image was annotated by two expert radiologists following two different approaches to delineate the lungs (as discussed in \cite{anatomicalmasks}). The first approach was to segment only the air cavity of the lung field, i.e. segmenting only the dark areas (regions of lucency) and ignoring areas of increased attenuation (opacities), which correspond to infected lung tissue. Following \cite{larrazabal2019anatomical} we call them \emph{air} masks.
In the second approach, annotators delineated the expected anatomy of the lung, including opaque areas following a comparative approach, mirroring the normal lung field onto the abnormal one. We call these \emph{anatomy} masks. 

\begin{figure}
    \centering
    \includegraphics[width=\linewidth]{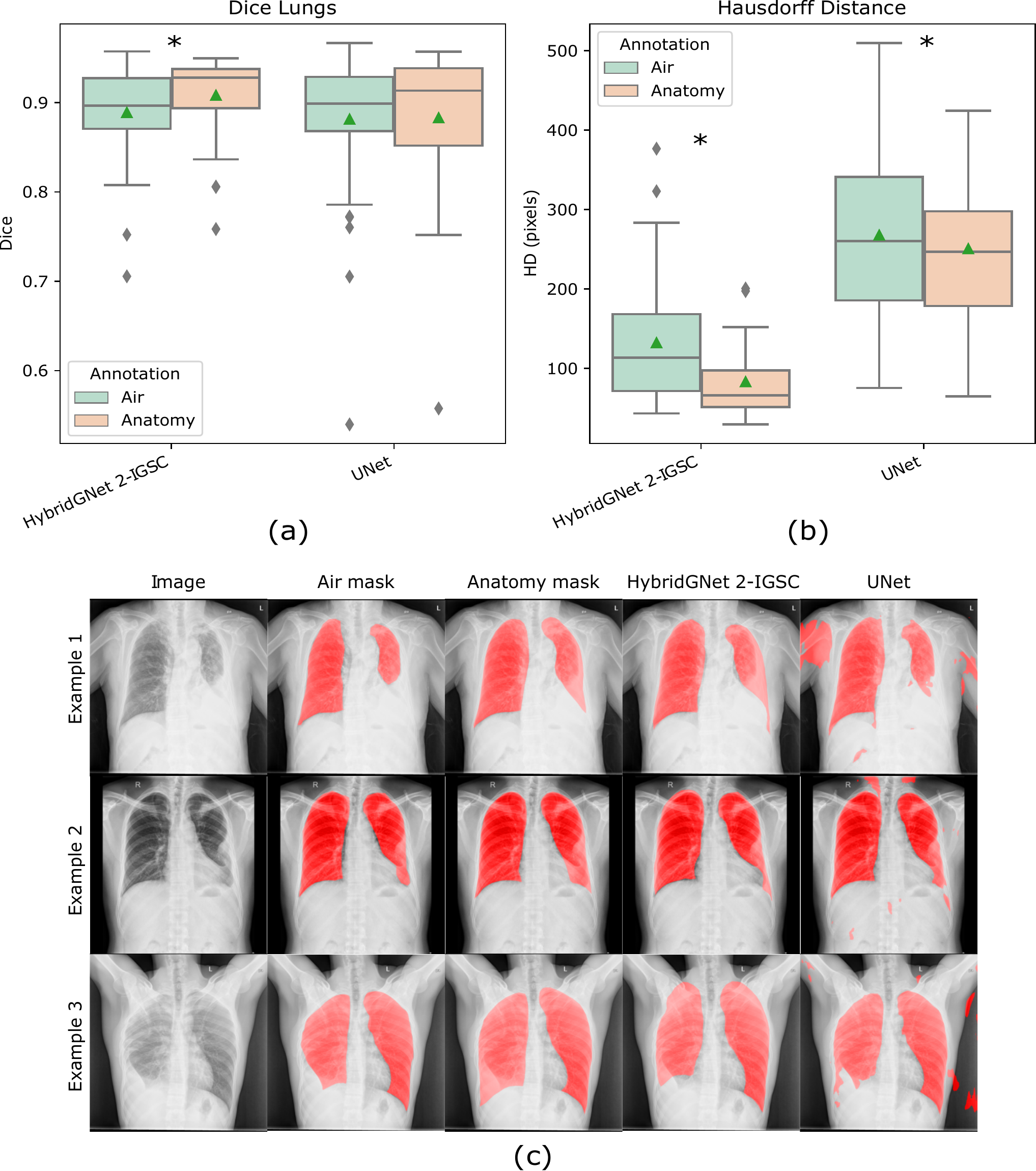}
    \caption{\noindent \textbf{Model behaviour on pathological anatomy.} Top boxplots show Dice coefficient (a) and HD (b), taking the air mask as ground truth (green) and taking the anatomy mask as ground truth (orange) for the HybridGNet and UNet models. * indicates significant differences between means according to Wilcoxon's test (p-value $<$ 0.05). 
    (c) Visual examples for the air and anatomy masks, and outputs given by both models. Results shows that our model tends to predict masks that follow the expected shape of the organs, while UNet predictions resamble the visible air section of the lungs.}
    \label{fig:air_anato}
\end{figure}

We compared the segmentation performance of HybridGNet and UNet considering both types of annotations as ground-truth, when trained on JSRT (which contains only masks of non-pathological lungs). 
Quantitative results shown in Figures \ref{fig:air_anato} (a) and (b) confirm that HybridGNet obtains results that are much closer to the \emph{anatomy} masks than to the \emph{air} masks, obtaining a higher Dice coefficient and a lower HD. Wilcoxon's test showed that the difference between the means on both metrics was indeed significative. Conversely, this tendency is less pronounced for UNet predictions, suggesting that HybridGNet encourages more anatomically plausible predictions, while UNet focuses on local texture patterns.
Figure \ref{fig:air_anato} (c) shows examples of air and anatomy masks, and qualitative results of both methods for three different images.
Regarding clinical utility, this opens the door for applications that combine both architectures: for example, the severity of tuberculosis infection could be estimated by measuring the difference between the UNet mask, representing the non-infected lung regions, and the HybdriGNet mask, representing the healthy lung area if there was no lung collapse.

\subsubsection{Cardiothoracic ratio estimation clinical use-case}
\label{sec:ctr}
A relevant intended use for lung and heart segmentation in chest x-rays is the detection of heart diseases, by identifying an enlargement of the cardiac silhouette. In radiology, this is done by measuring the CTR on a posteroanterior chest x-ray. This is calculated as the ratio between the (maximal) horizontal diameters of the heart and the thorax (inner edge of ribs/edge of pleura), which are manually measured by radiologists \cite{ctr}. A normal CTR lays between $0.42$ and $0.50$, while a CTR$\ >0.5$ is considered an abnormal finding. For example, in young patients it might indicate a heart disease, such as cardiomegaly or pericardial effusion. 

Manual calculation of CTR introduces observer variation and it is time consuming. Thus, here we evaluated the performance of HybridGNet for CTR estimation using a testing subset of 100 images from Padchest: 50 images with a \emph{cardiomegaly} label, and 50 without this label. Two radiology specialists from Hospital Italiano de Buenos Aires collaborated in our study by manually calculating the CTR for this subset. The mean CTR among them was considered as ground-truth. To reduce the DS due to the change of medical center and the lack of pathological anatomy, we constructed an augmented training set by merging the JSRT images with a subset of 117 images from Padchest with \emph{cardiomegaly} label.
Since Padchest did not originally include landmark annotations, in this augmented set we used the ones generated from dense segmentations in the experiment described in section \ref{sec:generatingLandmarks}.

\begin{figure}[t]
    \centering
    \includegraphics[width=\linewidth]{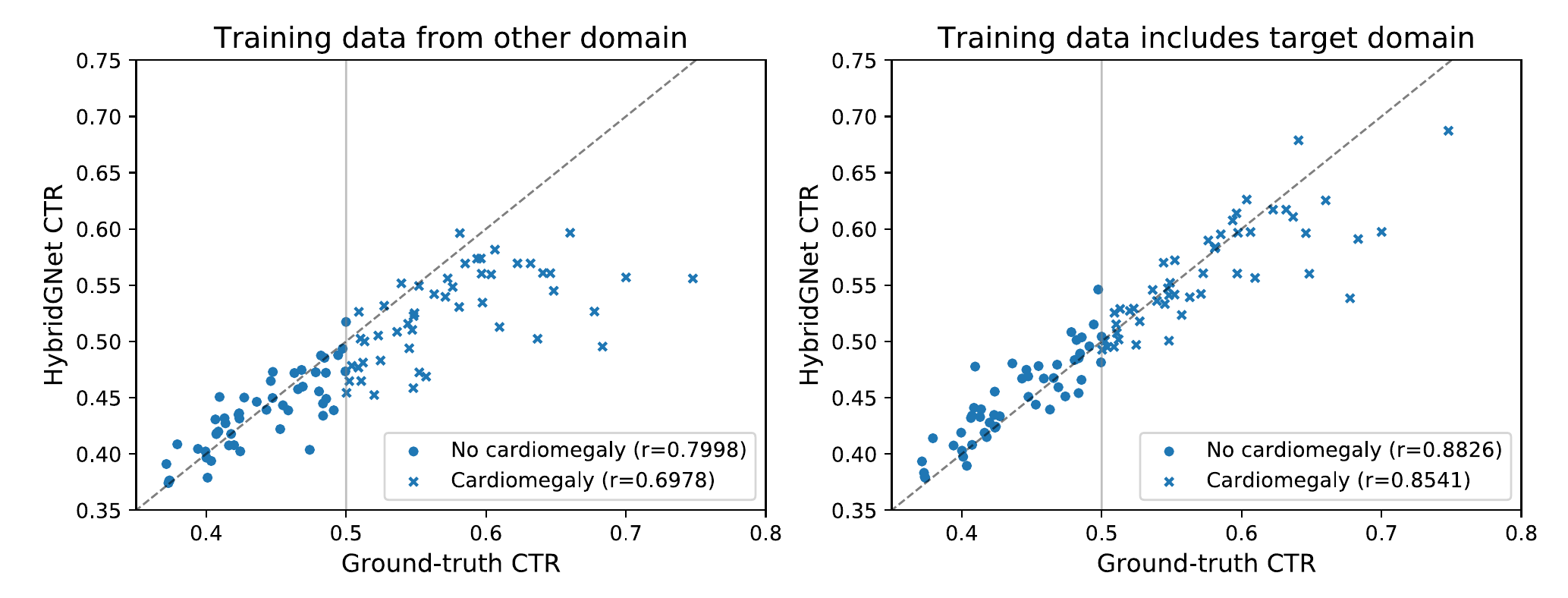}
    \caption{\noindent \textbf{CTR study.}
    Ground-truth CTR vs HybridGNet CTR when training with JSRT dataset only (left) and the dataset augmented with cardiomegaly images from Padchest (right). The vertical line indicates the boundary between normal and abnormal CTR.}
    \label{fig:ict}
\end{figure}

We compared model performance when training solely with JSRT images and when training with the augmented dataset. The predicted CTR was calculated automatically from HybridGNet outputs by measuring the maximum horizontal distance between lung borders and the maximum horizontal diameter of the heart mask. We found that the Pearson correlation coefficient between ground-truth CTR and predicted CTR increased when the model was trained with the augmented dataset. For the images with a ground-truth CTR$\ <0.5$ (normal cardiac silhouette) correlation increased from $0.80$ to $0.88$ when target-domain images where included during training. This improvement was even stronger for abnormal cases (CTR$\ >0.5$), increasing from $0.70$ to $0.85$. Figure \ref{fig:ict} shows a scatter plot of the 100 test images as data points, where the diagonal represents a perfect agreement between the CTR measurement of HybridGNet and physicians. We can see how the model trained solely with normal cardiac silhouette cases (JSRT) tends to underestimate the CTR, while the model trained with target-domain cases improves CTR calculation on abnormal hearts. These results suggest that even when using models which encourage anatomically plausibility, the construction of diverse databases (i.e. including representative samples of the target population) is still needed so that performance is maintained in real clinical scenarios.

\section{Conclusions}
In this paper we introduced HybridGNet, a new method to perform landmark based anatomical segmentation via hybrid graph neural networks with image-to-graph localized skip connections. Our study confirms that incorporating connectivity information through the graph adjacency matrix helps to improve anatomical plausibility and accuracy of the results when compared with other landmark-based and pixel-level segmentation models.
We also showcased several application scenarios for HybridGNet in the context of chest-x ray image analysis, and assessed its robustness with respect to different types of domain shift and image occlusions. When compared with dense pixel-level prediction models, we observed that HybridGNet achieves faster inference time, is much more robust to strong image occlusions and produces more anatomically plausible results in these contexts. We also evaluated the clinical utility of our model in the context of cardiothoracic ratio estimation and audited potential biases that may appear due to under-representation of certain demographic groups or pathologies. Our results go in line with the evidence reported in recent studies on fairness in biomedical image segmentation, highlighting the importance of constructing diverse databases which include representative demographic samples from the targeted population. In the future, we plan to extend the proposed HybridGNet model to volumetric images, where graphs can be used to represent meshes instead of contours.

\section{Acknowledgements}

We thank Alexandros Karargyris, Sema Candemir and Stefan Jaeger for sharing the segmentation masks used in the pathological anatomy experiment. We also thank Facundo Diaz and Martina Aineseder -specialists from the Radiology Department at Hospital Italiano de Buenos Aires- for their collaboration in the annotation of the Padchest images.

\bibliography{bib.bib}
\bibliographystyle{IEEEtran}

\end{document}

%% file: table_JSRT.tex

\begin{table*}[]
\centering
\caption{Landmark-based anatomical segmentation results for JSRT dataset. Mean (std). HD in millimeters.}
\begin{tabular}{r}
\hline
\bf Type\\
\hline 
\\ \\ \\ \\
\vspace{0.4pt}
\\ Landmark
\\ Methods
\\ \\
\vspace{0.4pt}
\\ \\ \\
\hline 
Pixel-level\\
Methods\\ \\
\hline
\end{tabular}%
\begin{tabular}{ll r r r r r}
\hline
\multicolumn{2}{l}{{\bf Model}} & {\bf MSE\he\he\he} & {\bf Dice Lungs\hh\hh} & {\bf HD Lungs\he} & {\bf Dice Heart\hf} & {\bf HD Heart\he} \\
\hline 
\multicolumn{2}{l}{PCA} & 340.024 (243.549) & 0.945 (0.014) & 17.445 \hf(9.669) & 0.906 (0.037) & 14.602 \hf(5.400) \\ 
\multicolumn{2}{l}{FC} & 332.197 (242.379) & 0.945 (0.017) & 17.535 (10.352) & 0.910 (0.038) & 15.020 \hf(5.785) \\ 
\multicolumn{2}{l}{MultiAtlas} & 492.262 (298.138) & 0.944 (0.013) & 20.317 \hf(9.344) & 0.886 (0.056) & 16.780 \hf(6.839) \\ 
\multicolumn{2}{l}{HybridGNet (without IGSC)} & 294.621 (274.497) & 0.952 (0.013) & 15.642 (10.922) & 0.913 (0.038) & 13.658 \hf(5.548) \\
\hline 
\multicolumn{1}{c}{} & Layer 3 & 277.536 (298.725) & 0.954 (0.014) & 14.565 (11.441) & 0.917 (0.037) & 13.401 \hf(5.376) \\ 
\multicolumn{1}{c}{} & Layer 4 & 288.597 (272.538) & 0.956 (0.013) & 16.054 (11.284) & 0.916 (0.038) & 14.153 \hf(6.038) \\ 
\multicolumn{1}{c}{} & Layer 5 & 258.413 (245.724) & 0.963 (0.010) & 13.662 (11.107) & 0.915 (0.039) & 13.738 \hf(5.181) \\ 
\multicolumn{1}{c}{\multirow{-4}{*}{1 IGSC}} & Layer 6 & 250.123 (232.032) & 0.960 (0.011) & 14.378 \hf(9.262) & 0.924 (0.030) & {12.339} \hf(4.844) \\
\hline 
\multicolumn{1}{c}{} & Layers 4-3 & 263.973 (262.700) & 0.963 (0.011) & 14.942 (10.589) & 0.921 (0.036) & 13.198 \hf(5.514) \\ 
\multicolumn{1}{c}{} & Layers 5-4 & 246.845 (230.235) & 0.968 (0.009) & 13.692 (10.984) & 0.924 (0.040) & 13.417 \hf(6.144) \\  
\multicolumn{1}{c}{\multirow{-3}{*}{2 IGSC}} & Layers 6-5 & {\bf 200.748} (211.080) & {\bf 0.974} (0.007) & {\bf 12.089} \hf(9.344) & {\bf 0.933} (0.031) & {\bf 11.613} \hf(5.581) \\
\hline 
\multicolumn{2}{l}{UNet} & $-$ \hspace{24px} & {0.981} (0.008) & 21.839 (26.291) & {0.942} (0.030) & 25.176 (34.570) \\
\multicolumn{2}{l}{UNet + Post-DAE} & $-$ \hspace{24px} & {0.965} (0.010) & 17.969 (14.457) & {0.935} (0.029) & 15.444 (14.283) \\
\multicolumn{2}{l}{nnUNet} & $-$ \hspace{24px} & {\bf 0.984} (0.005) & {\bf 9.615} \hf (7.874) & {\bf 0.952} (0.023) & {\bf 9.782} \hf (5.006) \\

\hline
\end{tabular}

\label{results-jsrt}
\end{table*}

%% file: table_inference_time.tex

\begin{table}[]
\centering
\caption{Inference times for each model per input image.}
\begin{tabular}{rlrl}
\hline 
\multicolumn{2}{r}{\textbf{Model}} & \multicolumn{2}{r}{\textbf{Time (s)}} \\
\hline 
\multicolumn{2}{r}{HybridGNet 2-IGSC}      & \multicolumn{2}{r}{\bf{0.53}} \\
\multicolumn{2}{r}{UNet}                  & \multicolumn{2}{r}{0.60} \\
\multicolumn{2}{r}{UNet + Post-DAE}        & \multicolumn{2}{r}{1.54} \\
\multicolumn{2}{r}{nnUNet}        & \multicolumn{2}{r}{2.76} \\
\hline 
\end{tabular}
\label{inference_times}
\end{table}

%% file: table_S2G.tex

\begin{table}[]
\centering
\caption{Results for generating landmark annotations from dense segmentations in the jsrt dataset. Mean (std). HD in millimeters.}

{\scriptsize
\begin{tabular}{l@{ }r@{\he}r@{\he}r@{\he}r@{\he}r@{\he}r}
\hline
{\bf Model} & {\bf MSE\he\hf} & {\bf Dice Lungs\hh\hh} & {\bf HD Lungs} & {\bf Dice Heart} & {\bf HD Heart} \\
\hline
PCA 
&    77.2\hh(133.7)&     0.978\hh(0.009)&    6.02\hh(3.46)&     0.97\hh(0.007)&     4.37\hh(1.61) \\
FC
&   105.3\hh(173.2)&     0.970\hh(0.014)&    7.82\hh(3.96)&     0.96\hh(0.014)&     5.78\hh(2.94) \\
Multi-atlas
&   236.3\hh(244.8)&{\bf0.991}\hh(0.004)&   10.98\hh(8.53)& {\bf0.99}\hh(0.006)&     4.64\hh(2.48) \\
HybridGNet
&    96.9\hh(145.0)&     0.970\hh(0.009)&    7.65\hh(3.75)&     0.96\hh(0.013)&     6.02\hh(2.77) \\
1 IGSC: L6
&    70.5\hh(144.9)&     0.983\hh(0.005)&    5.54\hh(5.30)&     0.97\hh(0.011)&     4.02\hh(2.24) \\
2 IGSC: L6-5 
&{\bf55.1}\hh(113.4)&{\bf0.991}\hh(0.003)& {\bf3.92}\hh(4.42)&{\bf0.99}\hh(0.005)& {\bf2.58}\hh(1.59) \\
\hline
\end{tabular}}

\label{tables2g}
\end{table}

%% file: table_Montgomery_Shenzhen.tex

\begin{table}[]
\centering
\caption{Domain shift results for landmark-based anatomical segmentation from JSRT dataset to Montgomery and Shenzhen. Mean (std). HD in pixels.}

{\scriptsize
\begin{tabular}{l@{ }r@{\he}r@{\he\he}r@{\he}r@{\he}}
\hline
\multirow{3}{*}{{\bf Model}}
& \multicolumn{2}{c}{{\bf Montgomery}} & \multicolumn{2}{c}{{\bf Shenzhen}} \\ \cline{2-5} 
& {\bf Dice Lungs\hh\hh} & {\bf HD Lungs\hf} & {\bf Dice Lungs\hf} & {\bf HD Lungs\he} \\ 
\hline
PCA 
& 0.906 \hh(0.082) & 60.08 \hh(36.89) & 0.894 \hh(0.054) & 79.12 \hf(47.73) \\
FC 
& 0.897 \hh(0.087) & 60.02 \hh(35.77) & 0.895 \hh(0.051) & 77.11 \hf(48.15) \\
Multi-alas 
& 0.909 \hh(0.080) & 61.77 \hh(31.62) & 0.900 \hh(0.054) & 88.13 \hf(48.94) \\
HybridGNet 
& 0.909 \hh(0.070) & 55.97 \hh(35.70) & 0.901 \hh(0.047) & 72.13 \hf(47.40) \\
1 IGSC: L6 
& 0.930 \hh(0.062) & 48.22 \hh(33.43) & 0.914 \hh(0.044) & 67.39 \hf(48.53) \\
2 IGSC: L6-5 
& {\bf0.954} \hh(0.043) & {\bf45.50} \hh(32.48) & {\bf0.935} \hh(0.038) & {\bf64.46} \hf(51.53) \\
\hline
UNet
& 0.944 \hh(0.068) & 127.72 \hh(97.76) & 0.933 \hh(0.055) & 220.89 \hh(102.94) \\
UNet+PostDAE
& 0.907 \hh(0.102) & 119.53 \hh(85.10) & 0.906 \hh(0.083) & 135.05 \hf(97.32) \\
nnUNet
& {\bf0.955} \hh(0.068) & {\bf44.79} \hh(60.31) & {\bf0.949} \hh(0.055) & {\bf61.31} \hf(67.97) \\
\hline
\end{tabular}}

\label{results-mont-shen}
\end{table}